\documentclass[a4paper,11pt]{article}
\usepackage[utf8]{inputenc}
\usepackage[T1]{fontenc}
\usepackage{textcomp}
\usepackage{fancyhdr}
\usepackage{fullpage}
\usepackage{lmodern}
\usepackage{amsmath,amssymb,bm,bbm}
\usepackage{algorithm}
\usepackage{algorithmic}
\usepackage{float}
\usepackage{graphicx}
\usepackage{extarrows}
\usepackage{caption}
\usepackage{graphicx, subfig}
\usepackage{titling}
\usepackage{yhmath}
\usepackage{mathrsfs}
\usepackage{cite}
\usepackage{footnote}
\usepackage{amsthm}
\usepackage{color}
\usepackage{slashed}
\usepackage{verbatim}
\usepackage[colorlinks=true,linkcolor=red,citecolor=blue]{hyperref}
\usepackage{amsfonts,euscript}
\usepackage{latexsym, multicol}
\usepackage{epstopdf}
\usepackage{psfrag}
\usepackage{mathrsfs}
\usepackage{xcolor}
\usepackage{comment}
\usepackage{authblk}
\usepackage{indentfirst}
\usepackage{lipsum}
\DeclareFontFamily{U}{mathx}{\hyphenchar\font45}
\DeclareFontShape{U}{mathx}{m}{n}{
<5> <6> <7> <8> <9> <10>
<10.95> <12> <14.4> <17.28> <20.74> <24.88>
mathx10}{}

\DeclareSymbolFont{mathx}{U}{mathx}{m}{n}
\DeclareFontSubstitution{U}{mathx}{m}{n}
\DeclareMathAccent{\widecheck}{0}{mathx}{"71}
\numberwithin{equation}{section}
\allowdisplaybreaks[3]

\def\etabf{\bm{\eta}}

\def\SSS{\mathbb{S}}
\def\Ve{{}^eV}
\def\Vi{{}^iV}
\def\Vie{{^i_eV}}
\def\Vii{{^i_iV}}
\def\ujp{\langle u \rangle}
\def\RRR{\mathbb{R}}
\def\slg{\slashed{g}}

\def\kah{\widehat{\ka}}

\def\Vphi{\varPhi}
\def\nabs{\slashed{\nab}}
\def\sdivs{\slashed{\div}}

\def\ka{\kappa}

\def\eps{\epsilon}
\def\pih{\widehat{\pi}}
\renewcommand{\c}{\cdot}

\newcommand{\pa}{\partial}
\newcommand{\pr}{\pa}
\newcommand{\unu}{{\underline{u}}}
\newcommand{\unc}{{\underline{C}}}
\newcommand{\bg}{\mathbf{g}}

\newcommand{\unchi}{\underline{\chi}}
\newcommand{\omb}{\underline{\omega}}
\newcommand{\ue}{\underline{\eta}}
\newcommand{\sld}{\slashed{d}}

\def\sfr{\mathfrak{s}}
\def\sk{\sfr}
\def\NNN{\mathbb{N}}

\newcommand{\mr}{\mathcal{R}}

\newcommand{\M}{\mathcal{M}}
\newcommand{\D}{\mathbf{D}}

\newcommand{\cuv}{{C_u^V}}
\newcommand{\ucuv}{{\unc_\unu^V}}

\newcommand{\Ric}{{\mathbf{Ric}}}
\DeclareMathOperator{\sRic}{Ric}
\newcommand{\g}{\bg}
\newcommand{\R}{{\mathbf{R}}}

\renewcommand{\div}{\sdiv}

\DeclareMathOperator{\lot}{l.o.t.}

\def\MM{\mathcal{M}}
\def\ub{\unu}
\def\chib{\unchi}
\def\Cb{\unc}
\def\om{\omega}
\def\ze{\zeta}
\def\Om{\Omega}
\def\etab{\ue}
\def\aa{\underline{\a}}
\def\bb{\underline{\b}}
\def\Si{\Sigma}
\def\si{\sigma}

\def\ga{\gamma}
\def\Ga{\Gamma}
\def\xib{\underline{\xi}}
\def\a{\alpha}
\def\b{\beta}

\def\trch{\tr\chi}
\def\trchb{\tr\chib}

\def\ka{\kappa}

\def\de{\delta}
\def\De{\Delta}
\def\nab{\nabla}

\def\ep{\varepsilon}
\def\les{\lesssim}
\def\RR{\mr}

\DeclareMathOperator{\curl}{curl}

\DeclareMathOperator{\tr}{tr}
\DeclareMathOperator{\sdiv}{div}
\def\bdiv{\mathbf{Div}}
\DeclareMathOperator{\ric}{Ric}

\newtheorem{thm}{Theorem}[section]
\newtheorem{prop}[thm]{Proposition}
\newtheorem{lem}[thm]{Lemma}

\newtheorem{rk}[thm]{Remark}
\newtheorem{df}[thm]{Definition}

\title{The Stability of Minkowski Spacetime}
\author{Dawei Shen}
\begin{document}
\maketitle
\begin{abstract}
The nonlinear stability of Minkowski spacetime has been one of the central achievements in the mathematical theory of general relativity and, more broadly, in the analysis of nonlinear geometric wave equations. Since the seminal work of Christodoulou-Klainerman, the problem has shaped fundamental advances in our understanding of decay, dispersion, and the intricate interplay between geometry and analysis in the Einstein vacuum equations.

This survey presents an overview of the main ideas and techniques underlying the stability theory of Minkowski spacetime. We emphasize the role of decay assumptions, geometric foliations, energy identities, and gauge choices in the global analysis. Particular attention is devoted to exterior stability results, minimal decay regimes, and the borderline case, where the failure of spacetime integrability for nonlinear interactions reveals subtle threshold phenomena. 

Our goal is to provide a coherent perspective on the evolution of the field, to clarify the structural mechanisms behind the known results, and to outline some of the central open problems that remain in the borderline regime.
\end{abstract}
\tableofcontents
\section{Introduction}
\subsection{Einstein vacuum equations and the Cauchy problem}
A Lorentzian $4$--manifold $(\MM,\g)$ is called a vacuum spacetime if it satisfies the Einstein vacuum equations
\begin{equation}\label{EVE}
    \Ric(\g)=0 \qquad \text{ in }\; \MM,
\end{equation}
where $\Ric$ denotes the Ricci tensor of the Lorentzian metric $\g$. Due to the invariance of \eqref{EVE} under diffeomorphisms, solutions are naturally considered up to coordinate transformations. In general coordinates, the Einstein vacuum equations form a nonlinear, geometric, second-order system of partial differential equations for the metric $\g$. In suitable gauges, for instance, in wave coordinates, the system can be shown to be hyperbolic, which allows one to formulate a well-posed initial value problem.

The Cauchy problem for the Einstein vacuum equations is posed by prescribing an initial data set $(\Si,g,k)$, where $(\Si,g)$ is a Riemannian $3$--manifold and $k$ is a symmetric $2$--tensor on $\Sigma$ satisfying the constraint equations
\begin{align}
\begin{split}\label{constraintk}
    R=|k|^2-(\tr k)^2, \qquad\quad\nab^j k_{ij}=\nab_i \tr k,
\end{split}
\end{align}
where $R$ denotes the scalar curvature of $g$ and $\nab$ its Levi-Civita connection. Here
\[
|k|^2:=g^{im}g^{jl}k_{ij}k_{lm},\qquad\quad\tr k:=g^{ij}k_{ij}.
\]
Given a solution $(\MM,\g)$ that arises as the future development of such data, the hypersurface $\Sigma\subset\MM$ is spacelike, with induced metric $g$ and the second fundamental form $k$.

The foundational well-posedness theory for the Einstein vacuum equations, developed in \cite{cb,cbg}, guaranties that for any smooth initial data set $(\Si,g,k)$ satisfying \eqref{constraintk}, there exists a unique smooth maximal global hyperbolic development $(\M,\g)$ solving \eqref{EVE}, in which $\Si$ is embedded as a Cauchy hypersurface with induced data $(g,k)$.

The most basic example of a vacuum spacetime is Minkowski space $(\M,\etabf)$, given by
\[
\MM=\mathbb{R}^4,\qquad\quad\etabf=-(dt)^2+(dx^1)^2+(dx^2)^2+(dx^3)^2,
\]
whose associated initial data are
\[
\Si=\mathbb{R}^3, \qquad\quad g=(dx^1)^2+(dx^2)^2+(dx^3)^2, \qquad\quad k=0.
\]
The central theme of this survey is the stability of Minkowski spacetime under perturbations of its initial data. We begin by reviewing the current state of the art on this problem and the main ideas underlying the known stability results.
\subsection{Overview of previous works}
The global nonlinear stability of Minkowski spacetime for the Einstein vacuum equations was first established by Christodoulou and Klainerman in their seminal work \cite{Ch-Kl}. This result, obtained within a maximal foliation framework, introduced a fundamentally geometric approach combining energy estimates based on the Bel-Robinson tensor with a refined vectorfield method. It marked a major milestone in mathematical general relativity and laid the foundation for much of the subsequent work on nonlinear stability problems.

Building on this framework, Bieri \cite{Bieri,Bieri2009} provided an alternative proof of global stability under weaker regularity and decay assumptions. Her approach still relies on a maximal foliation, but requires fewer derivatives and a reduced set of commutation vectorfields, thereby clarifying the robustness of the geometric method introduced in \cite{Ch-Kl}.

Several complementary approaches have since been developed, exploring different gauges and geometric structures. Klainerman and Nicol\`o \cite{Kl-Ni} proved the stability of Minkowski spacetime in the exterior of an outgoing null cone using a double null foliation, highlighting the role of characteristic geometry in stability problems. Under stronger assumptions on asymptotic decay and regularity, they further established sharp peeling estimates for curvature components \cite{knpeeling}, elucidating the asymptotic structure of solutions arising from asymptotically flat initial data.

A different perspective was introduced by Lindblad and Rodnianski in \cite{lr1,lr2}, who proved the stability of Minkowski spacetime in wave coordinates. Their analysis revealed that the Einstein equations exhibit a \emph{weak null structure} in this gauge, allowing one to treat the problem within a quasilinear wave equation framework. This wave-coordinate approach was further developed by Huneau \cite{huneau}, who proved nonlinear stability results using generalized wave coordinates adapted to translation Killing fields. Using tools from microlocal analysis and Melrose’s $b$--calculus, Hintz and Vasy \cite{hv} reproved the stability of Minkowski spacetime, offering a scattering-theoretic interpretation of the long-time dynamics. Graf \cite{graf} established global nonlinear stability in the context of the spacelike--characteristic Cauchy problem, which, when combined with the exterior analysis of \cite{Kl-Ni}, provides an alternative route to recovering the original result of \cite{Ch-Kl}.

Within the double null framework of \cite{Kl-Ni}, and exploiting the $r^p$--weighted energy estimates introduced by Dafermos and Rodnianski \cite{Da-Ro}, exterior stability results were revisited and refined. In particular, the exterior stability of Minkowski spacetime was reproved by the author in \cite{Shen22}, emphasizing the effectiveness of $r^p$--weighted methods in capturing decay along null directions and in controlling exterior regions. More recently, Hintz \cite{Hintz}, using the microlocal framework developed in \cite{hv}, also reproved the exterior stability of Minkowski spacetime.

Beyond the vacuum setting, the stability of Minkowski spacetime has also been established for Einstein’s equations coupled to various matter models. For the Einstein--Maxwell system, Zipser \cite{zipser} extended the geometric framework of \cite{Ch-Kl} to prove nonlinear stability. Using wave-coordinate methods, Loizelet \cite{lo09} studied the Einstein--scalar field--Maxwell system in $(1+n)$ dimensions, $n\geq 3$, while Speck \cite{speck} proved global nonlinear stability for a family of electromagnetic field models in $(1+3)$ dimensions.

For the Einstein-Klein-Gordon system, Lefloch and Ma \cite{leflochma15,lefloch} and Wang \cite{wang} independently proved global stability for initial data coinciding with the Schwarzschild solution outside a compact set. Ionescu and Pausader \cite{ionescu} subsequently established global stability for general small initial data, while Lefloch and Ma \cite{leflochMa} introduced a foliation combining hyperboloidal and constant-time slices to treat related problems.

Stability results have also been obtained for the Einstein--Vlasov system. Taylor \cite{taylor} studied the massless case with compact support on the mass shell, while Fajman, Joudioux, and Smulevici \cite{fajman} treated the massive case with Schwarzschild asymptotics in the exterior region. Lindblad and Taylor \cite{lt} considered massive Vlasov matter with compact spatial support, whereas Bigorgne, Fajman, Joudioux, Smulevici, and Thaller \cite{bigo} addressed the massless case for general initial data. Finally, Wang \cite{wxc} established stability results for the massive Einstein--Vlasov system with general small initial data.

We also mention recent progress on the Einstein-Yang-Mills system, including exterior stability in $(1+3)$ dimensions \cite{Ghanem2310.08196} and global stability in higher dimensions \cite{Ghanem2310.07954}.

For a broader survey of stability results for the Einstein equations coupled with various matter models, we refer to Section 5.4 of \cite{smulevici}.

Taken together, these works illustrate the diversity of methods—geometric, hyperbolic, microlocal, and dispersive—that have been brought to bear on the stability problem of Minkowski spacetime, and underscore its role as a central testing ground for techniques in nonlinear wave equations and geometric analysis.

These results demonstrate that the stability of Minkowski spacetime is not a phenomenon restricted to the vacuum case, but rather reflects a robust structural property of the Einstein equations under small perturbations.
\subsection{Asymptotic flatness}
We recall several notions that are intrinsic to the Christodoulou-Klainerman framework and are necessary for the formulation of the main stability theorem. We begin with the geometric setting of the initial data. In \cite{Ch-Kl}, an analysis is performed for initial data posed on maximal hypersurfaces.
\begin{df}\label{def6.1}
An initial data set $(\Si,g,k)$ is said to be posed on a maximal hypersurface if
\begin{equation}
    \tr k=0.
\end{equation}
In this case, $(\Si,g,k)$ is called a maximal initial data set, and the Einstein constraint equations \eqref{constraintk} reduce to
\begin{equation}
    R=|k|^2,\qquad \sdiv k=0,\qquad\tr k=0.
\end{equation}
\end{df}
To quantify both the decay and the regularity of the initial data, Christodoulou-Klainerman \cite{Ch-Kl} introduce a notion of asymptotic flatness formulated in terms of weighted asymptotics for the induced metric $g$ and the second fundamental form $k$.
\begin{df}\label{def6.3}
Let $s\geq 1$ and $q\in\mathbb{N}$. An initial data set $(\Sigma_0,g,k)$ is called $(s,q)$--asymptotically flat if there exists a coordinate system $(x^1,x^2,x^3)$ defined outside a sufficiently large compact set such that the following conditions hold:
\begin{itemize}
\item If $s\geq 3$\footnote{The notation $f=o(1)$ means that $f\to 0$ as $r\to\infty$. Moreover, for $m\in\mathbb{R}$ and $q\in\mathbb{N}$, we write $f=o_q(r^{-m})$ if $\partial^\alpha f=o(r^{-m-|\alpha|})$ for all $|\alpha|\le q$.}
\begin{align}
\begin{split}\label{old1.3}
g_{ij}&=\left(1-\frac{2M}{r}\right)^{-1}dr^2+r^2 d\sigma_{\SSS^2}+o_{q+1}\bigl(r^{-\frac{s-1}{2}}\bigr),\\
k_{ij}&=o_q\bigl(r^{-\frac{s+1}{2}}\bigr).
\end{split}
\end{align}
\item If $1\le s<3$,
\begin{align}
\begin{split}\label{gks1}
g_{ij}&=\delta_{ij}+o_{q+1}\bigl(r^{-\frac{s-1}{2}}\bigr),\\
k_{ij}&=o_q\bigl(r^{-\frac{s+1}{2}}\bigr).
\end{split}
\end{align}
\end{itemize}
\end{df}
An important and largely independent question concerns the existence of $(s,q)$--asymptotically flat initial data sets for prescribed values of $(s,q)$. While Christodoulou-Klainerman assume strong asymptotic flatness in order to establish stability, subsequent works have investigated to what extent such assumptions are necessary and how flexible the class of admissible initial data can be.

This direction was initiated by Carlotto-Schoen \cite{CS}, who achieved conic localization of vacuum initial data at the expense of a decay loss in the transition region. Carlotto later conjectured that such localization should be possible without loss of decay \cite[Open Problem~3.18]{Car21}. This conjecture was resolved by Aretakis-Czimek-Rodnianski \cite{ACR} via characteristic gluing.

Mao-Tao \cite{MaoTao} subsequently revisited the conic setting and developed a spacelike construction based on explicit solution operators for the linearized constraint equations with prescribed conic support; see also \cite{MOT} for solution operators with annulus support. On the other hand, Fang-Szeftel-Touati \cite{FST} constructed $(s,q)$--asymptotically flat vacuum initial data for $s>1$ and $q\in\NNN$ sufficiently large using the conformal method. More recently, \cite{SW2602} established the existence of $(s,q)$--asymptotically flat initial data for all $s\in[1,3)$ and $q\in\NNN$ sufficiently large. It is also worth mentioning the recent elliptic-transport construction of spacelike initial data adapted to a sphere foliation due to Chen-Klainerman \cite{ChenKlainerman}.

To measure the smallness of the initial data in a scale-invariant way, Christodoulou-Klainerman introduced weighted energy-type functionals adapted to the decay parameter $s$. We briefly recall these quantities.
\begin{df}\label{dffunctional}
Let $d_0$ denote the geodesic distance to a fixed point $O\in\Si_0$. Let $B_{ij}:=(\curl \widehat{R})_{ij}$ be the Bach tensor, where $\widehat{R}$ denotes the traceless part of the Ricci tensor. For $q\geq 2$, define:
\begin{itemize}
\item If $s\geq 3$,
\begin{align}
\begin{split}\label{Ckfunctional}
J_0^{(q)}(\Sigma_0,g,k)&:=\sup_{\Si_0}\Bigl((d_0^2+1)^3|\ric|^2\Bigr)+\sum_{l=0}^q\int_{\Si_0}(d_0^2+1)^{l+\frac{s-2}{2}}|\nab^l k|^2 \\
&\quad+\sum_{l=0}^{q-2}\int_{\Si_0}(d_0^2+1)^{l+\frac{s+2}{2}}|\nab^l B|^2.
\end{split}
\end{align}
\item If $1\leq s<3$,
\begin{align}
\begin{split}\label{bierifunctional}
J_0^{(q)}(\Si_0,g,k):=\sum_{l=0}^q\int_{\Si_0}(d_0^2+1)^{l+\frac{s-2}{2}}|\nab^l k|^2+\sum_{l=0}^{q-1}\int_{\Si_0}(d_0^2+1)^{l+\frac{s}{2}}|\nab^l\ric|^2.
\end{split}
\end{align}
\end{itemize}
\end{df}
\subsection{ADM energy and momentum}
An important aspect of the study of asymptotically flat spacetimes is the behavior of the ADM energy, linear momentum, and angular momentum. In the classical setting of Christodoulou-Klainerman \cite{Ch-Kl}, the strong decay assumptions on the initial data ensure that these quantities are well-defined and finite.

More precisely, let $S_r\subset\Si_0$ denote the coordinate sphere of radius $r$, with outward unit normal $N$ and induced area element $d\si$. The ADM energy is defined by
\begin{align*}
E := \frac{1}{16\pi} \lim_{r\to\infty}\int_{S_r}\left(\pr_i g_{ij} - \pr_j g_{ii} \right) N^j \, d\sigma,
\end{align*}
the linear momentum by
\begin{align*}
P_i := \frac{1}{8\pi}\lim_{r \to \infty} \int_{S_r}\left(k_{ij}-(\tr k)\, g_{ij}\right) N^j \, d\sigma, \qquad i = 1,2,3,
\end{align*}
and the angular momentum by
\begin{align*}
J_l:= \frac{1}{8\pi}\lim_{r \to \infty} \int_{S_r}{\in_{li}}^{j} x^i \left(k_{jm}-(\tr k)\, g_{jm}\right) N^m\, d\si, \qquad l = 1,2,3.
\end{align*}
The existence of these limits depends crucially on the decay rate of the initial data. Under the asymptotic assumptions in Definition \ref{old1.3}, the ADM energy $E$ and linear momentum $P$ are well-defined provided that $s>2$, while the angular momentum $J$ requires the stronger condition $s>3$. See Chapter 3 of \cite{Ch} for a detailed discussion. In particular, the assumptions in \cite{Ch-Kl} correspond to $(4,3)$--asymptotically flat initial data in Definition \ref{old1.3}. Thus, all three quantities are finite and characterize the total energy, linear momentum, and angular momentum of the spacetime as measured at infinity.

In contrast, in the weaker decay regimes considered in Bieri’s extension \cite{Bieri, Bieri2009} and in the author's works \cite{Shen23, Shen24}, the assumptions correspond to $s=2$, $s\in(1,2]$ and $s=1$, respectively. In these cases, the decay is no longer sufficient to guarantee the convergence of the above integrals, and the ADM energy, linear momentum, and angular momentum are, in general, not well-defined. This reflects the fact that these frameworks allow for a broader class of asymptotically flat spacetimes. From a physical perspective, this indicates that such spacetimes cannot, in general, be interpreted as isolated systems with finite total conserved quantities.
\subsection{Minkowski stability in \texorpdfstring{\cite{Ch-Kl}}{}}\label{ssec6.2}
In this subsection, we review the seminal work of Christodoulou-Klainerman \cite{Ch-Kl} on the nonlinear stability of Minkowski spacetime. This result provides the first complete proof of global nonlinear stability for the Einstein vacuum equations and has served as a foundational reference for all subsequent developments in this area. The proof is formulated within a geometric framework based on maximal foliations and combines refined energy estimates, vectorfield methods, and the Bel-Robinson tensor in an essential way. We can now state the main stability theorem of Christodoulou-Klainerman.
\begin{thm}[Christodoulou-Klainerman \cite{Ch-Kl}]\label{ckmain}
There exists $\ep_0>0$ sufficiently small such that if the initial data set $(\Si_0,g,k)$ is maximal, $(4,3)$--asymptotically flat, and satisfies
\[
J_0^{(3)}(\Si_0,g,k)\leq\ep_0^2,
\]
then its maximal globally hyperbolic development is unique, smooth, geodesically complete, and globally asymptotically flat. Moreover, there exist a global maximal time function $t$ and an optical function $u$ defined in an exterior region.
\end{thm}
In \cite{Ch-Kl}, the initial data is assumed to be strongly asymptotically flat, reflecting the high level of decay and regularity required by the original argument. Since then, a variety of alternative asymptotic regimes and decay assumptions have been explored in the literature on the stability of Minkowski spacetime; we refer to Section~\ref{sec:Shen} for a discussion of these developments.
\section{The Einstein vacuum equations in wave coordinates}
In this section, we briefly recall the formulation of the Einstein vacuum equations in wave coordinates. This reduction plays a central role in analytic approaches to the stability of Minkowski spacetime, as it allows one to recast the Einstein equations as a system of quasilinear wave equations. Throughout this section, we restrict ourselves to the vacuum case.
\subsection{Wave coordinates and the reduced Einstein equations}
Let $(\MM,\g)$ be a Lorentzian $4$--manifold. The Einstein vacuum equations are given by
\begin{equation}\label{EVE-wave}
\Ric(\g)=0.
\end{equation}
These equations are invariant under diffeomorphisms, which reflects the gauge freedom inherent in General Relativity. In order to obtain a hyperbolic formulation suitable for analysis, one must fix this gauge freedom.

A natural and widely used choice is the \emph{wave coordinate gauge}. A coordinate system $(x^\a)_{\a=0}^3$ is said to be a wave coordinate system if each coordinate function satisfies the covariant wave equation
\begin{equation}\label{wave-gauge}
\square_\g x^\a=0,
\end{equation}
where $\square_\g := \g^{\mu\nu}\D_\mu\D_\nu$ denotes the Laplace-Beltrami operator associated with $\g$. Equivalently, the wave coordinate condition can be written as
\begin{equation}\label{wave-gauge2}
\Ga^\a:= \g^{\mu\nu}\Ga^\a_{\mu\nu}=0,
\end{equation}
where $\Ga^\a_{\mu\nu}$ are the Christoffel symbols of $\g$.

Under the wave coordinate condition \eqref{wave-gauge2}, the Einstein vacuum equations reduce to a system of quasilinear wave equations for the metric components $\g_{\alpha\beta}$. More precisely, one obtains the \emph{reduced Einstein equations}
\begin{equation}\label{reduced-Einstein}
\square_\g\g_{\a\b} = F_{\a\b}(\g,\pr\g),
\end{equation}
where $F_{\a\b}$ is quadratic in the first derivatives of $\g$.

The nonlinear terms $F_{\a\b}$ appearing in \eqref{reduced-Einstein} can be written schematically as
\begin{equation}\label{nonlinear-structure}
F_{\a\b}(\g,\pr\g)=\g^{\mu\nu}\g^{\rho\sigma}\bigl(\pr_\mu\g_{\a\rho}\,\partial_\nu \g_{\beta\sigma}+\pr_\mu \g_{\beta\rho}\,\partial_\nu \g_{\alpha\sigma}-\pr_\mu\g_{\rho\sigma}\,\pr_\nu\g_{\a\b}\bigr)+\lot,
\end{equation}
where $\lot$ denotes lower order terms depending on $\g$ and $\pr\g$.

In particular, the reduced Einstein equations form a system of quasilinear wave equations whose quadratic nonlinearities do \emph{not} satisfy the classical null condition. This failure was already observed in early work by Choquet-Bruhat \cite{cb} and constitutes a major obstruction to the application of standard small-data global existence results for wave equations in three spatial dimensions.
\subsection{Propagation of the wave gauge}
An important feature of the wave coordinate formulation is that the gauge condition propagates dynamically. More precisely, if the initial data on a Cauchy hypersurface satisfy
\begin{equation}\label{gauge-initial}
\Ga^\a\big|_{\Si_0}=0,\qquad\pr_t\Ga^\a\big|_{\Si_0}=0,
\end{equation}
then the solution of the reduced system \eqref{reduced-Einstein} automatically satisfies
\[
\Ga^\a \equiv 0
\]
throughout the domain of dependence.

As a consequence, solutions of the reduced Einstein equations \eqref{reduced-Einstein} which satisfy the wave gauge condition on the initial hypersurface are, in fact, solutions of the full Einstein vacuum equations \eqref{EVE-wave}.
\subsection{Perturbations of Minkowski spacetime}
To study the stability of Minkowski spacetime, one writes the metric as a perturbation of the Minkowski metric $\eta$:
\begin{equation}\label{metric-perturb}
\g_{\a\b} =\etabf_{\a\b}+h_{\a\b}.
\end{equation}
In wave coordinates, the reduced Einstein equations \eqref{reduced-Einstein} can be rewritten schematically as
\begin{equation}\label{reduced-h}
\square_{\etabf} h_{\a\b}=Q_{\a\b}(\pr h,\pr h)+\text{ higher order terms},
\end{equation}
where $\square_{\etabf}$ denotes the flat wave operator and $Q_{\alpha\beta}$ is quadratic in the first derivatives of $h$.

This formulation makes clear that the stability of Minkowski spacetime in wave coordinates reduces to a small-data global existence problem for a system of quasilinear wave equations on $\RRR^{1+3}$. However, because of the absence of the classical null condition, additional structural properties are required to control the long-time behavior of solutions.
\subsection{The weak null structure}
Although the quadratic nonlinearities in \eqref{reduced-h} do not satisfy the classical null condition, Lindblad and Rodnianski \cite{lr1,lr2} observed that the reduced Einstein equations enjoy a weaker structural property, now known as the \emph{weak null condition}.

Roughly speaking, this condition asserts that the most slowly decaying components of the solution do not interact in a way that leads to uncontrollable growth. Although certain metric components may exhibit mild growth along outgoing null directions, this growth does not feed back into the full system and remains compatible with global existence. This can be understood by analyzing the associated asymptotic system along outgoing null cones.

This observation forms the basis for the wave-coordinate proof of the nonlinear stability of Minkowski spacetime and provides a complementary analytical framework to the geometric approach of Christodoulou-Klainerman \cite{Ch-Kl}. See Section \ref{sec3} for more explanations. In contrast, the geometric approach of Christodoulou–Klainerman, which we review in Sections \ref{sec:CK} and \ref{sec:Shen}, incorporates these structures directly at the level of curvature and null geometry.
\section{Quasilinear wave equations and the vectorfield method}\label{sec3}
In view of the reduced Einstein equations in wave gauge, which take the form of quasilinear wave equations, it is natural to begin the analysis of Minkowski stability by recalling general strategies for global existence for small-data quasilinear wave equations. This section provides a brief overview of the vectorfield method, the null condition, and related structural properties that play a fundamental role in the analysis of Einstein equations.
\subsection{A general strategy for quasilinear wave equations}
We consider systems of quasilinear wave equations of the form
\begin{equation}\label{eq:qlwave}
(-\pr_t^2+\De)u^i + F^{\a\b}(u)\pr_\a\pr_\b u^i=Q^i(\pr u,\pr u),
\end{equation}
with initial data
\begin{equation*}
u(0,x)=u_0(x),\qquad \partial_t u(0,x)=u_1(x),
\end{equation*}
where $u=(u^i)_{1\le i\le N}$ is an $\RRR^N$--valued function, the coefficients $F^{\a\b}$ are smooth and symmetric in $\a,\b$, vanishing at $u=0$, and $Q^i$ are quadratic forms in the first derivatives of $u$.

The standard global existence strategy relies on the combination of:
\begin{itemize}
\item a hierarchy of weighted energy norms controlling derivatives of $u$,
\item energy inequalities propagating these norms in time,
\item decay estimates allowing control of the nonlinear terms,
\item and a bootstrap argument closing the estimates globally in time.
\end{itemize}
Denoting by $E_N[u(t)]$ the top-order energy norm controlling up to $N$ derivatives, one seeks estimates of the form
\begin{equation}\label{eq:energy-ineq}
E_N[u(t)] \le C E_N[u(0)] + \int_0^t \mathcal{N}(s)\,ds,
\end{equation}
where $\mathcal{N}(s)$ represents nonlinear error terms. If one can establish decay estimates such as
\begin{equation}\label{eq:decay-nl}
|\mathcal{N}(s)|\les E_N[u(s)]\langle s\rangle^{-q},
\end{equation}
with $q>1$, then Gr\"onwall-type arguments yield global-in-time bounds for sufficiently small initial data.
\subsection{The vectorfield method of Klainerman}
A key tool for deriving decay estimates for solutions of the wave equation is the vectorfield method introduced by Klainerman. Consider the standard wave operator
\[
\square:=-\pr_t^2+\De
\]
on $\mathbb{R}^{1+n}$. We introduce the collection $\mathcal{Z}$ of vectorfields consisting of:
\begin{itemize}
\item translations $\pr_t,\pr_{x_i}$,
\item spatial rotations $\Om_{ij}=x_i\pr_{x_j}-x_j\pr_{x_i}$,
\item Lorentz boosts $\Om_{0i}=t\pr_{x_i}+x_i\pr_t$,
\item the scaling vectorfield $S=t\pr_t+\sum_i x_i\pr_{x_i}$.
\end{itemize}
These vectorfields satisfy the commutation relations
\[
[\square,Z]=0,\qquad Z\in\mathcal{Z},
\]
except for the scaling vectorfield, for which $[\square,S]=2\square$. Using multi-index notation $Z^\alpha$, one defines the energy norm
\[
E_N[u(t)] := \sum_{|\alpha|\le N} \int_{\mathbb{R}^n}\bigl(|\partial_t Z^\alpha u|^2 + |\nabla Z^\alpha u|^2\bigr)\,dx.
\]
A central ingredient is the Klainerman-Sobolev inequality, which yields pointwise decay estimates from weighted $L^2$ bounds. In particular, for sufficiently regular $\psi$,
\begin{equation}\label{eq:KS}
|\psi(t,x)| \lesssim \frac{1}{(1+t+|x|)^{\frac{n-1}{2}}(1+|t-|x||)^{\frac12}}
\sum_{|\alpha|\le \frac{n+2}{2}} \|Z^\alpha\psi(t,\cdot)\|_{L^2_x}.
\end{equation}
Applying this inequality to derivatives of solutions of the linear wave equation yields sharp decay estimates away from the light cone.
\subsection{The role of dimension and the null condition}
In spatial dimension $n\ge4$, the decay rate obtained from \eqref{eq:KS} is sufficient to ensure $q>1$ in \eqref{eq:decay-nl}, leading to global existence for small data under general assumptions on the nonlinearities.

In contrast, in dimension $n=3$, the borderline decay rate leads only to $q=1$, which is insufficient to close the bootstrap argument in general. This obstruction is reflected in classical blow-up results for quasilinear wave equations. The borderline nature of three spatial dimensions is precisely what makes the stability of Minkowski spacetime such a delicate problem.

Global existence in three spatial dimensions can nevertheless be obtained when the nonlinearities satisfy the \emph{null condition}. Roughly speaking, this condition requires that quadratic nonlinear terms vanish when evaluated on null directions, ensuring that at least one derivative in each quadratic interaction is tangential to the light cone and thus enjoys improved decay.
\subsection{The Einstein equations in wave gauge and the weak null condition}
When written in wave coordinates, the Einstein vacuum equations reduce to a system of quasilinear wave equations of the form
\begin{equation}\label{eq:einstein-wave}
\square_\g \g_{\alpha\beta} = Q_{\alpha\beta}(\partial \g,\partial \g),
\end{equation}
where the quadratic terms $Q_{\alpha\beta}$ do not satisfy the classical null condition. This failure was already observed by Choquet-Bruhat \cite{cb} and rules out a direct application of the standard small-data theory.

Nevertheless, Lindblad and Rodnianski \cite{lr1,lr2} identified a weaker structural property, known as the \emph{weak null condition}, which captures cancellations arising from the geometric structure of the equations and the wave gauge constraints. While certain components of the metric may exhibit mild (typically logarithmic) growth, this growth does not propagate through the entire system, allowing for global existence and decay estimates.

This observation led to an alternative proof of the nonlinear stability of Minkowski spacetime based on wave coordinates, providing a complementary analytical framework to the geometric approach of Christodoulou-Klainerman. \cite{Ch-Kl}.
\subsection{Remarks on geometric versus analytic approaches}
The wave-coordinate approach offers considerable flexibility and is well adapted to analytic techniques, including Fourier and microlocal analysis. However, geometric properties such as curvature decay, the structure of null infinity, or the Bondi mass loss law require additional geometric input beyond the global existence in wave coordinates.

By contrast, the geometric framework developed by Christodoulou-Klainerman in \cite{Ch-Kl} encodes these features directly into the analysis at the cost of substantial technical complexity. Both perspectives have since played a central role in subsequent developments and extensions of Minkowski stability.
\section{Stability of Minkowski in \texorpdfstring{\cite{Ch-Kl}}{}}\label{sec:CK}
This section gives a brief overview of the proof of Christodoulou-Klainerman \cite{Ch-Kl}. 
The argument combines a global geometric gauge (a maximal foliation), a null decomposition adapted to an optical function, and a vectorfield method based on the Bel-Robinson tensor. 
A central theme is that different null components of the curvature obey different decay rates, and the proof exploits a hierarchy of weighted energy estimates consistent with this structure.
\subsection{Geometric set-up}\label{ssec7.1}
\subsubsection{Maximal foliation and the \texorpdfstring{$3+1$}{} decomposition}\label{maximalfoliation}
The spacetime $(\mathcal{M},\g)$ is foliated by spacelike hypersurfaces $\Si_t$ given as level sets of a time function $t$. Let $T$ be the future normal unit to $\Si_t$ and write the metric in ADM form
\begin{equation}\label{metricSi_survey}
\g=-\phi^2 dt^2 + g_{ij}\,dx^i dx^j,
\end{equation}
where $\phi$ is the lapse, characterized by $T=\phi \nabla t$. 
The second fundamental form of the slices is
\begin{equation}\label{dfk_survey}
k_{ij}:=-\g(\nabla_i T,\partial_j).
\end{equation}
Christodoulou-Klainerman work in the \emph{maximal gauge} $\tr k=0$, which simplifies both the constraints on $\Sigma_t$ and the evolution system. In particular, in vacuum one obtains the maximal constraints
\begin{equation}\label{max_constraints_survey}
\tr k=0,\qquad\quad \nabla^j k_{ji}=0,\qquad\quad R=|k|^2,
\end{equation}
together with the lapse equation
\begin{equation}\label{lapseequation_survey}
\Delta \phi=\phi |k|^2,
\end{equation}
and the evolution equations for $(g,k)$, see (1.0.11)--(1.0.13) in \cite{Ch-Kl}.
\subsubsection{Maximal-null foliation and null frames}\label{nullfoliation}
To capture the propagation along null directions and to formulate the decay hierarchy, one introduces an outgoing optical function $u$ solving the eikonal equation
\begin{equation}\label{eikonal_survey}
\g^{\mu\nu}\partial_\mu u\,\partial_\nu u=0,
\end{equation}
and considers the associated null hypersurfaces $C_u=\{u=\text{const}\}$. 
Intersecting $C_u$ with $\Si_t$ yields a foliation by topological spheres $S(t,u)=\Sigma_t\cap C_u$. 
On $\mathcal{M}\setminus \Phi$ (away from the axis), one constructs a null frame $(e_1,e_2,e_3,e_4)$ adapted to this double foliation, with $e_A$ tangent to $S(t,u)$ and $e_3,e_4$ incoming/outgoing null directions.

Relative to this null frame, one uses the standard null decomposition of the Ricci coefficients and the curvature as follows:
\begin{align}
\begin{split}\label{defga}
\chib_{AB}&=\g(\D_A e_3, e_B),\qquad\quad \chi_{AB}=\g(\D_A e_4, e_B),\\
\xib_A&=\frac 1 2 \g(\D_3 e_3,e_A),\qquad\quad\,\,  \xi_A=\frac 1 2 \g(\D_4 e_4, e_A),\\
\omb&=\frac 1 4 \g(\D_3e_3 ,e_4),\qquad\quad \,\,\,\;\, \om=\frac 1 4 \g(\D_4 e_4, e_3), \\
\etab_A&=\frac 1 2 \g(\D_4 e_3, e_A),\qquad\quad\;  \eta_A=\frac 1 2 \g(\D_3 e_4, e_A),\\
\ze_A&=\frac 1 2 \g(\D_{e_A}e_4, e_3),
\end{split}
\end{align}
and
\begin{align}
\begin{split}\label{defr}
\a_{AB} &=\R(e_A, e_4, e_B, e_4),\qquad\quad\,\aa_{AB} =\R(e_A, e_3, e_B, e_3), \\
\b_{A} &=\frac 1 2\R(e_A, e_4, e_3, e_4),\qquad\quad\,\,\bb_{A}=\frac 1 2 \R(e_A, e_3, e_3, e_4),\\
\rho&=\frac 1 4 \R(e_3, e_4, e_3, e_4), \qquad\quad\;\;\;\; \si =\frac{1}{4}{^*\R}(e_3, e_4, e_3,e_4),
\end{split}
\end{align}
where $^*\R$ denotes the Hodge dual of $\R$. This decomposition is the starting point for the curvature estimates and for the bookkeeping of error terms. This foliation is the geometric backbone of the entire argument.
\subsection{Bel-Robinson tensor}\label{ssec:BR}
In this subsection, we introduce the Bel-Robinson tensor, which serves as the analog of
the energy-momentum tensor for the gravitational field.

Let $(\MM,\g)$ be a vacuum spacetime and let $(e_1,e_2,e_3,e_4)$ be a null frame,
where $e_3$ and $e_4$ are respectively incoming and outgoing null vectors,
normalized so that
\[
\g(e_3,e_4)=-2,
\]
and $(e_1,e_2)$ form an orthonormal basis tangent to the $2$--spheres
$S_{t,u}$.
These components encode the propagation of curvature along null directions. In particular, $\a$ and $\aa$ represent the purely outgoing and incoming gravitational radiation, respectively, while $(\b,\rho,\si,\bb)$ correspond to intermediate components with mixed null behavior.
\subsubsection{The Bel-Robinson tensor}
In vacuum, the Weyl tensor satisfies the Bianchi equations
\[
\D^\a W_{\a\b\ga\de}=0.
\]
Associated to $W$ is the Bel-Robinson tensor $Q$, defined by
\begin{equation}\label{def:BR}
Q_{\alpha\beta\gamma\delta}
:= W_{\alpha\mu\beta\nu} W_{\gamma}{}^{\mu}{}_{\delta}{}^{\nu}
+ {}^\star W_{\alpha\mu\beta\nu}\,{}^\star W_{\gamma}{}^{\mu}{}_{\delta}{}^{\nu}.
\end{equation}

The Bel-Robinson tensor enjoys several key properties which make it suitable
for energy estimates:
\begin{itemize}
\item $Q$ is completely symmetric and tracefree;
\item $Q$ is divergence-free in vacuum:
\[
\nabla^\alpha Q_{\alpha\beta\gamma\delta}=0;
\]
\item for any future-directed timelike vector $X$, the quantity
\[
Q(X,X,X,X)\ge 0,
\]
so that $Q$ defines a positive energy density.
\end{itemize}
In this sense, the Bel-Robinson tensor plays the role of an energy-momentum tensor
for the gravitational field. Contracting $Q$ against suitable vectorfields
and applying the divergence theorem yields conserved or almost-conserved
energy quantities for the Weyl curvature.
\subsubsection{Energy densities in null components}
When contracted against combinations of timelike and null vectorfields,
the Bel-Robinson tensor produces weighted sums of the null curvature components.
For instance, relative to the standard null frame, one finds schematically
\[
Q(T,T,T,T)\sim|\a|^2+|\b|^2+|\rho|^2+|\si|^2+|\bb|^2+|\aa|^2,
\]
where $T$ is a timelike unit normal to the maximal hypersurfaces.

More refined contractions, involving null or conformal vectorfields,
assign different weights to different curvature components, reflecting
their distinct decay and propagation properties. This feature is essential
in capturing the hierarchy of decay rates needed in the nonlinear stability
analysis and will be exploited in the vectorfield method discussed in the following.
\subsection{The vectorfield method and the conformal Morawetz vectorfield}\label{ssec:K0}
The vectorfield method provides a systematic mechanism for extracting
energy and decay estimates from the Bel-Robinson tensor.
In the Christodoulou-Klainerman framework, suitable multipliers are applied
to the Weyl curvature in order to generate spacetime energy identities,
which ultimately yield both coercive control and integrated decay.
A central role is played by the conformal Morawetz vectorfield $K_0$.
\subsubsection{Bel-Robinson energy currents and multipliers.}
Let $Q$ denote the Bel-Robinson tensor associated to the Weyl curvature $W$.
Given vectorfields $X,Y,Z$, one defines the Bel-Robinson energy current
\begin{equation}\label{BRcurrent}
P_\alpha[W;X,Y,Z]:=Q_{\alpha\beta\gamma\delta}X^\beta Y^\gamma Z^\delta.
\end{equation}
In vacuum, using the divergence-free property of $Q$, one computes
\begin{equation}\label{BRdiv}
\D^\a P_\a[W;X,Y,Z]
=
Q_{\a\b\ga\de}
\left(
\D^{(\a}X^{\b)}Y^\ga Z^\de
+
\D^{(\a}Y^{\b)}X^\ga Z^\de
+
\D^{(\a}Z^{\b)}X^\ga Y^\de
\right),
\end{equation}
where parentheses denote symmetrization.

Spacetime energy identities are obtained by integrating
\eqref{BRdiv} over a spacetime region and applying the divergence theorem.
The deformation tensors of the vectorfields $X,Y,Z$
determine the bulk error terms and hence govern the structure of the estimates.

The simplest multiplier choice consists in taking all three vectorfields equal
to the future-directed unit normal $T$ to the maximal hypersurfaces $\Sigma_t$.
The associated energy
\[
\mathcal{E}(t):=\int_{\Sigma_t} Q(T,T,T,T)
\]
controls the $L^2$--norm of all null curvature components on $\Sigma_t$.
This energy is coercive and provides uniform control of curvature on spacelike slices,
but by itself it does not yield spacetime decay.

To extract dispersive decay, one must employ vectorfields
adapted to the geometry of Minkowski spacetime.
\subsubsection{The conformal Morawetz vectorfield \texorpdfstring{$K_0$}{} and weighted spacetime control.}
A key ingredient in the work of Christodoulou-Klainerman
is the conformal Morawetz vectorfield
\begin{equation}\label{defK0}
K_0:=\frac12\bigl(u^2 e_3+\underline{u}^2 e_4\bigr),
\end{equation}
where $u$ and $\ub$ denote the outgoing and incoming optical functions,
and $(e_3,e_4)$ are the associated null generators.

In Minkowski spacetime, $K_0$ is a conformal Killing vectorfield.
In the perturbed setting, its deformation tensor generates error terms,
but these are of lower order and can be controlled under the bootstrap assumptions.

Contracting the Bel-Robinson tensor with $(T,T,K_0)$ yields the energy current
\[
P_\alpha[W;T,T,K_0]
=
Q_{\alpha\beta\gamma\delta}T^\beta T^\gamma K_0^\delta.
\]
Applying the divergence theorem in a spacetime region $\mathcal{D}$
bounded by hypersurfaces $\Sigma_{t_1}$, $\Sigma_{t_2}$ and null boundaries,
one obtains an identity of the form
\begin{align}\label{BRK0identity}
\int_{\Sigma_{t_2}} Q(T,T,K_0,T)
+
\int_{\mathcal{D}} \mathcal{I}_{K_0}[W]
=
\int_{\Sigma_{t_1}} Q(T,T,K_0,T)
+
\int_{\mathcal{D}} \mathcal{E}_{K_0}[W],
\end{align}
where $\mathcal{I}_{K_0}[W]$ denotes the bulk spacetime term generated
by the deformation tensor of $K_0$, and $\mathcal{E}_{K_0}[W]$
collects error terms involving curvature and connection coefficients.

A crucial feature of $K_0$ is that the spacetime term
$\mathcal{I}_{K_0}[W]$ controls weighted spacetime integrals
of the null curvature components.
Schematically,
\[
\mathcal{I}_{K_0}[W]\sim u^2|\aa|^2+\ub^2|\a|^2+u^2|\bb|^2+\ub^2|\b|^2+u\ub\left(|\rho|^2+|\si|^2\right).
\]
These weights reflect the peeling hierarchy and encode the expected decay
rates of the different curvature components.
In particular, the $K_0$--energy identity yields integrated decay estimates
which cannot be obtained from the timelike multiplier alone.

The combination of the $T$--energy identity and the $K_0$--energy identity
forms the backbone of the curvature estimates in the Christodoulou-Klainerman
framework. The $T$--energy controls the size of curvature on spacelike slices,
while the $K_0$--energy provides spacetime decay and captures
the dispersive character of gravitational radiation.

These estimates are subsequently refined by commuting with suitable vectorfields
and by exploiting the null structure of the Bianchi equations,
ultimately leading to the closure of the bootstrap argument.
\subsection{Higher-order curvature estimates: commutation and error terms}\label{ssec_commutation_signature}

To obtain higher-order curvature bounds and improved decay rates, the Bianchi equations are commuted with suitable vectorfields adapted to the geometry.
For a vectorfield $X$, this is achieved using the modified Lie derivative $\widehat{\mathcal{L}}_X W$, which preserves the algebraic symmetries and tracelessness of the Weyl tensor:
\begin{equation}\label{modifiedLie_survey}
\widehat{\mathcal{L}}_X W
:= \mathcal{L}_X W - \frac12 \bigl({}^{(X)}\pi \cdot W\bigr)
- \frac38 \tr^{(X)}\pi\, W,
\qquad {}^{(X)}\pi_{\alpha\beta}:=\frac12 \mathcal{L}_X g_{\alpha\beta}.
\end{equation}
Commuting the Bianchi equations then produces inhomogeneous equations of the schematic form
\begin{equation}\label{commutedBianchi_survey}
D^\alpha(\widehat{\mathcal{L}}_T W)_{\alpha\beta\gamma\delta}
=J(T;W)_{\beta\gamma\delta},
\end{equation}
where the source term $J(T;W)$ is quadratic in the deformation tensor ${}^{(T)}\pi$ and the curvature components.

Applying divergence identities to the Bel-Robinson tensor $Q[\widehat{\mathcal{L}}_T W]$ and contracting with combinations of the vectorfields $T$ and $K$ yields spacetime error integrals.
Controlling these terms requires a refined use of the null structure of the equations together with the decay hierarchy of curvature components.

A systematic framework for organizing such error terms is provided by the \emph{principle of conservation of signature}.
Each null curvature component is assigned a signature measuring the imbalance between $e_4$ and $e_3$ directions (for instance, $s(\alpha)=+2$ and $s(\aa)=-2$).
This bookkeeping principle constrains the possible nonlinear interactions, distinguishing borderline contributions from those which are manifestly integrable, and plays a key role in closing the higher-order energy estimates.

\subsection{Bootstrap scheme and the last slice argument}\label{ssec_bootstrap_last}
The global stability argument is implemented through a bootstrap scheme on a time interval $[0,t^\ast]$, combining weighted energy estimates with pointwise decay obtained via Sobolev inequalities on the spheres $S(t,u)$.
A representative top-order exterior bootstrap assumption takes the form
\begin{equation}\label{bootstrap_example_survey}
\left\|(1+u^2)^2 r^q \nabs^q\a\right\|_{L^2(r\ge r_0/2)}
+\left\| r^{q+2} \nabs^q \aa \right\|_{L^2(r\ge r_0/2)}
\leq \varepsilon_0,
\qquad q=0,1,2,
\end{equation}
and the goal is to improve all such bounds, thereby closing the bootstrap.

A distinctive feature of the Christodoulou-Klainerman argument is the treatment of the optical function $u$.
Since the eikonal equation is a nonlinear transport equation, prescribing $u$ on an initial slice amounts to fixing a radial foliation which is simply transported to the future, with no a priori reason for the geometry of the spheres $S(t,u)$ to improve. Instead, in the exterior region, the optical function is initialized from the \emph{last slice} $\Sigma_{t^\ast}$.
This allows the foliation on $\Sigma_{t^\ast}$ to be chosen so as to satisfy the improved geometric bounds obtained from the bootstrap assumptions.

Then extend the existence interval and reconstruct the optical function $u$ in the last two slices, eventually proving the convergence of the optical functions ${}^{(t)}u$ as $t\to+\infty$.
This last slice argument is essential for reconciling the dynamical construction of the null foliation with the decay properties required for global stability.
\section{Stability of Minkowski in \texorpdfstring{\cite{Shen23,Shen24}}{}}\label{sec:Shen}
This section gives a brief overview of the proofs in \cite{Shen23,Shen24}, which extend the stability of Minkowski spacetime to minimal decay and borderline decay regimes. 
We begin by stating the main stability results.
\subsection{Stability of Minkowski with minimal decay and borderline decay}
\begin{thm}[Global stability of Minkowski with minimal decay \cite{Shen23}]\label{thm:Shen23}
Fix $s>1$ and $0<\ep_0\ll 1$. Let $(\Si_0,g,k)$ be a $(s,2)$--asymptotically flat vacuum initial data set of size $\ep_0$, in the sense of Definition \ref{def6.3}. Then $(\Si_0,g,k)$ admits a unique, globally hyperbolic, smooth, geodesically complete development under the Einstein vacuum equations. Moreover, this development is globally asymptotically flat, in the sense that the Riemann curvature tensor tends to zero along any future-directed causal or spacelike geodesic.
\end{thm}
Theorem \ref{thm:Shen23} shows that global nonlinear stability holds under the minimal decay assumption $s>1$. In other words, once one is slightly above the borderline threshold, small asymptotically flat initial data lead to complete and globally asymptotically flat development.

The situation changes in the critical case $s=1$. At present, global stability at this level of decay is not known. However, stability can still be established in the exterior region, as shown in the following result.
\begin{thm}[Exterior stability of Minkowski with borderline decay \cite{Shen24}]\label{thm:Shen24}
Let $(\Si_0,g,k)$ be a $(1,3)$--asymptotically flat \emph{vacuum} initial data set of size $\ep_0$, in the sense of Definition~\ref{def6.3}. 
Let $K\subset\Si_0$ be a compact set such that $\Si_0\setminus K$ is diffeomorphic to $\RR^3\setminus\overline{B}_1$. 
Then the stability of Minkowski spacetime holds in the exterior of an outgoing null cone. 
More precisely, there exists a unique future development $(\M,\g)$ of $\Si_0\setminus K$ with the following properties:
\begin{itemize}
\item $(\M,\g)$ admits a double null foliation $(C_u,\Cb_{\ub})$ whose outgoing leaves $C_u$ are complete;
\item all geometric and curvature quantities associated with the double null foliation satisfy quantitative control.
\end{itemize}
\end{thm}
\begin{rk}
Various choices of the parameters $s$ and $q$ in Definition~\ref{def6.3} have appeared in the literature, reflecting different decay and regularity regimes in which stability can be established. 
In particular:\footnote{This list focuses on low values of $s$. For larger values of $s$, see for example \cite{knpeeling} for $s>7$ and \cite{Shen22} for $s>3$.}
\begin{itemize}
\item $(s,q)=(7+\de,5)$ is used in \cite{knpeeling};\footnote{Here and in the lines below, $\de$ denotes a constant satisfying $0<\de\ll 1$.}
\item $(s,q)=(4,3)$ is used in \cite{Ch-Kl,Kl-Ni,graf};
\item $(s,q)=(3+\de,6)$ is used in \cite{lr2};
\item $(s,q)=(3+\de,26)$ is used in \cite{hv};
\item $(s,q)=(3+\de,2)$ is used in \cite{Shen22};
\item $(s,q)=(3+\de,2433)$ is used in \cite{Hintz};
\item $(s,q)=(3-\de,200)$ is used in \cite{ionescu} for the Einstein-Klein-Gordon system and in \cite{wxc} for the Einstein-Vlasov system;
\item $(s,q)=(2+\de,20)$ is used in \cite{leflochMa} for the Einstein-Klein-Gordon system;
\item $(s,q)=(2,2)$ is used in \cite{Bieri,Bieri2009}.
\end{itemize}
\end{rk}
Theorem~\ref{thm:Shen23} extends the global stability result of Christodoulou-Klainerman \cite{Ch-Kl} to the minimal decay regime $s>1$. 
More precisely, it establishes global nonlinear stability for $(s,2)$--asymptotically flat initial data with $s\in(1,2]$, thereby showing that the strong decay assumptions in \cite{Ch-Kl} are not essential for global control.

Theorem \ref{thm:Shen24} addresses the genuinely borderline case $(s,q)=(1,3)$. 
At this level of decay, global stability is no longer available with the current methods. 
Nevertheless, one can still prove exterior stability in the domain of dependence of an outgoing null cone. This result isolates the precise threshold at which global arguments break down while demonstrating that the dispersive mechanism remains effective in the exterior region.
\subsection{Maximal-null foliation}
As discussed in Section~\ref{sec:CK}, a central geometric ingredient in the Christodoulou-Klainerman framework is the construction of a foliation combining a maximal time function with an outgoing optical function. In \cite{Ch-Kl}, this maximal-null foliation is constructed from the \emph{last slice}, as described in Section \ref{ssec_bootstrap_last}. 
More precisely, the optical function is initialized on the final hypersurface $\Si_{t^\ast}$ in the bootstrap argument, so that the geometric properties of the spheres $S(t,u)$ can be improved at peak time and then propagated backward. This construction of the ``last slice'' is closely tied to the global bootstrap scheme and plays a decisive role in controlling the asymptotic geometry.

A fundamental difference in the analysis of \cite{Shen23}, compared to \cite{Ch-Kl,Bieri}, lies in the construction of the sphere foliation. Instead of initializing the optical function on a last slice, the null foliation is constructed directly from the \emph{initial slice} and a distinguished timelike curve, which plays the role of an axis of symmetry. This forward construction is more natural in the minimal decay regime and avoids relying on a posteriori improvements at the final time.

We now briefly describe the construction in \cite{Shen23}. We first proceed as in Section~\ref{maximalfoliation} to construct the maximal hypersurfaces $\Si_t$, and denote by $T$ the future-directed unit normal to $\Si_t$. 

Let $O\in\Si_0$ be a point such that the functional $J_0^{(2)}$ defined in Definition~\ref{dffunctional} is sufficiently small. 
Denote by $\Phi$ the integral curve of $O$ along the vectorfield $T$, which plays the role of a distinguished timelike curve and is referred to as the \emph{symmetry axis}. 

Starting from $O$, one constructs the outgoing null cone $C_0$ in the future of $\Sigma_0$. 
This cone separates the spacetime into two geometrically distinct regions. 
More precisely, the spacetime $\mathcal{M}$ admits the decomposition
\[
\mathcal{M}=I^+(O)\cup C_0\cup D^+(\Si_0\setminus O),
\]
where $I^+(O)$ denotes the future domain of influence of the point $O$, and $D^+(\Si_0\setminus O)$ denotes the future domain of dependence of $\Si_0\setminus O$. 
This decomposition provides the geometric framework for the subsequent construction of the null foliation.

We now describe the construction of the outgoing optical function $u$ on $\mathcal{M}$. 
Along the symmetry axis $\Phi$, $u$ is prescribed by
\[
T(u)=1\quad\text{ on }\Phi,\qquad u(O)=0.
\]
For each $p\in\Phi$, let $C_{u(p)}$ denote the outgoing null cone emanating from $p$. 
This yields a foliation of $I^+(O)$ by outgoing null hypersurfaces $C_u$ for $u>0$, and $u$ is extended to $I^+(O)$ by requiring that the hypersurfaces $C_u$ coincide with its level sets. 
We denote
\[
L:=-\nabla u \qquad \text{in } I^+(O).
\]
It remains to define $u$ in $D^+(\Si_0\setminus O)$.

To this end, define a scalar function $w$ on $\Si_0$ by $w=0$ at $O$ and extend it smoothly to $\Si_0$ so that $\nabla w\neq 0$ and $w\to\infty$ at spatial infinity. 
This fixes a radial foliation of the initial hypersurface by the level sets
\[
S_{(0)}(w_1)=\{p\in\Sigma_0:\ w(p)=w_1\},
\qquad w_1\ge 0.
\]
The optical function $u$ is then extended to $D^+(\Sigma_0\setminus O)$ as the solution to the eikonal equation
\begin{equation}\label{uSi0}
\g^{\mu\nu}\partial_\mu u\,\partial_\nu u=0,\qquad u|_{\Si_0\setminus O}=-w.
\end{equation}
Denoting
\[
L:=-\nabla u \qquad \text{in } D^+(\Sigma_0\setminus O),
\]
one has the geodesic equation $\nabla_L L=0$. 
Thus $u$ and $L$ are globally defined on $\mathcal{M}$, and $u$ is smooth by construction.

\begin{figure}
\centering
\includegraphics[width=15.2cm]{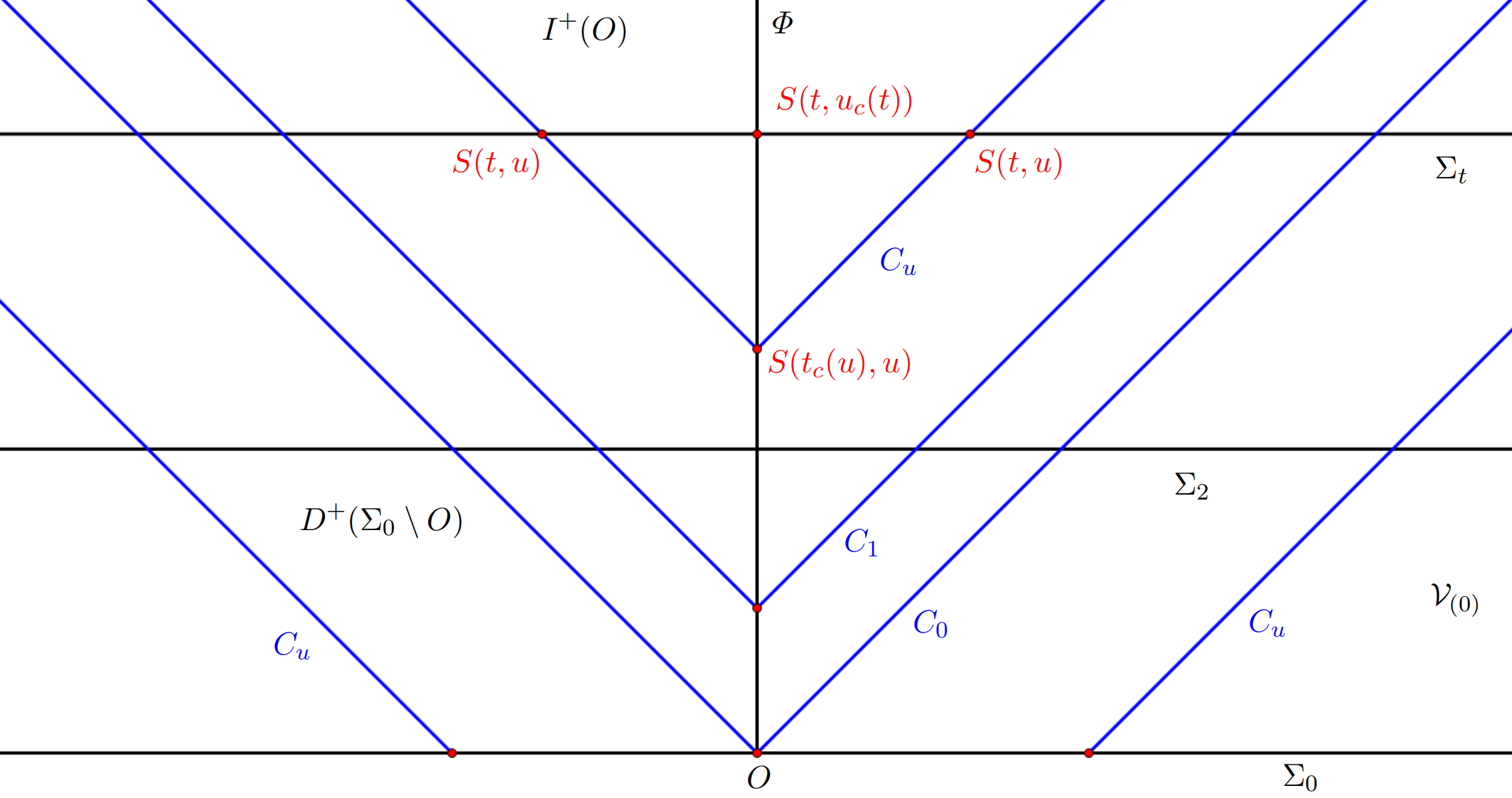}
\caption{Maximal-null foliation of the spacetime $\M$.}
\label{foliation}
\end{figure}

Points on the symmetry axis are viewed as spheres of radius $0$:
\begin{equation}\label{dfucttcu}
S(t,u_c(t)):=\Sigma_t\cap\Phi,\qquad 
S(t_c(u),u):=C_u\cap\Phi.
\end{equation}
Away from the axis, the spacetime is foliated by the $2$--spheres
\[
S(t,u):=\Sigma_t\cap C_u,\qquad u\in\mathbb{R},\quad t\ge 0.
\]
We note that
\[
u>0\ \text{in }I^+(O),\qquad 
u=0\ \text{on }C_0,\qquad 
u<0\ \text{in }D^+(\Sigma_0\setminus O),
\]
see Figure~\ref{foliation}.
We set
\[
V_t:=J^-(\Sigma_t)\cap J^+(\Sigma_2),
\]
and define the \emph{exterior} and \emph{interior} regions by
\[
^eV_t\{p\in V_t:\ u\le 1\},\qquad 
^iV_t:=\{p\in V_t:\ u\ge 1\}.
\]
Finally, we introduce the null frame adapted to this maximal-null foliation. 
Set
\[
a:=|\nabla u|^{-1},\qquad e_4:=aL.
\]
By the eikonal equation,
\[
0=g(\nabla u,\nabla u)=|\nabla u|^2-|T(u)|^2=a^{-2}-|T(u)|^2,
\]
hence $a=T(u)^{-1}$, and in particular $a=1$ on $\Phi$. 
Define
\[
N:=e_4-T,
\]
which is a vectorfield on $\mathcal{M}\setminus\Phi$. 
Then set $e_3:=T-N$; together with a local orthonormal frame $(e_1,e_2)$ tangent to $S(t,u)$, this produces the null frame $(e_1,e_2,e_3,e_4)$ used throughout. 
With respect to this null frame, one defines the Ricci coefficients and Weyl curvature components as in \eqref{defga} and \eqref{defr}.
\subsection{\texorpdfstring{$r^p$}{}--weighted estimates for curvature components}
The $r^p$--weighted energy method has become a central tool in the analysis of decay for geometric wave equations. It was first introduced by Dafermos-Rodnianski in \cite{Da-Ro} for linear wave equations on black hole backgrounds, with the aim of capturing decay at large radii without relying on the full vectorfield method. The method was subsequently adapted to the Bianchi equations by Holzegel in \cite{holzegel}.

In the setting of Minkowski stability, the $r^p$--weighted estimate method was first applied by the author in \cite{Shen22}, where it was used to reprove exterior stability within a double null framework. Later, in \cite{Shen23}, the method was further developed to treat not only the large $r$ region but also the regime where $r$ is near $0$, allowing one to obtain curvature estimates near the symmetry axis $\Vphi$.

This section describes the $r^p$--weighted energy method as applied to curvature components. The key idea is to exploit the null structure of the Bianchi equations in order to derive weighted divergence identities whose spacetime bulk terms encode the distinct decay rates of the various curvature components; see Proposition \ref{keyintegral} for more details.
\subsubsection{Comparison with the approaches of \texorpdfstring{\cite{Ch-Kl,Bieri}}{}.}
This section aims to clarify the differences between the approaches of \cite{Ch-Kl,Bieri}, as well as the role of the Bianchi equations in deriving the $r^p$--weighted estimates.

In the setting of \cite{Ch-Kl}, the analysis relies on strong asymptotic decay assumptions and high regularity of the initial data, leading to control of all curvature components via Klainerman’s invariant vector field method, as outlined in Section \ref{sec:CK}. Bieri \cite{Bieri,Bieri2009} relaxes these assumptions by allowing for weaker decay and lower regularity while preserving the essential structure of the energy estimates and reducing the number of commutation vectorfields required.

The approach in \cite{Shen23, Shen24} retains many of the structural features of these earlier works, in particular the use of the null structure of the Bianchi equations and the associated hierarchy of curvature components. However, a key difference lies in the formulation of the energy estimates. Instead of relying on a family of carefully constructed commutation vectorfields, the analysis is based on $r^p$--weighted estimates, which do not require the introduction of such vectorfields.

This leads to a conceptually simpler framework: the $r^p$--weighted method directly exploits the asymptotic behavior of spacetime without the need to track the commutation properties of vectorfields. In particular, it allows one to capture decay along null hypersurfaces in a more direct and flexible way.

In particular, while \cite{Ch-Kl,Bieri} relies on energy identities generated by the Bel-Robinson tensor and commutation vectorfields, the $r^p$--weighted method derives energy estimates directly from weighted divergence identities adapted to null directions. This leads to a different structure of the energy hierarchy in which decay is built into the weights rather than extracted through vectorfields.
\subsubsection{Estimates for Bianchi pairs}
Under the assumptions of Theorem \ref{thm:Shen23}, the Weyl curvature components satisfy the following weighted energy bound on the initial hypersurface $\Si_0$:
\begin{align}\label{initials}
\int_{\Si_0}r^s\left(|\a|^2+|\b|^2+|(\rho,\si)|^2+|\bb|^2+|\aa|^2\right)\les\ep_0^2, \qquad\quad s>1.
\end{align}
A central task in the analysis is to show that such energy bounds propagate into the future development of $\Si_0$. More precisely, one aims to establish corresponding flux estimates along null hypersurfaces. For instance, one seeks to prove that
\begin{align}
    \int_{C_u}r^s|\a|^2+\int_{\Cb_\ub}r^s|\b|^2\les\ep_0^2,
\end{align}
for any outgoing null cone $C_u$ and incoming null cone $\Cb_\ub$.

The derivation of these estimates relies on the Bianchi equations, which govern the evolution of the Weyl curvature. When expressed in a null frame, these equations decompose into a coupled system of transport and divergence-type equations for the various null components. Schematically, Bianchi equations take the following form:
\[
\nabs_4\b+2\trch\,\b \sim \sdivs\a+\lot,\qquad\nabs_3\a+\frac{1}{2}\trchb\,\a\sim \sld_2^*\b+\lot,
\]
together with analogous equations for the remaining components.

A key feature of the Bianchi equations is their null structure, which leads to favorable cancellations among tangential derivatives, see Lemma \ref{keypoint} for more details.

In the context of \cite{Shen22,Shen23,Shen24}, the Bianchi equations are combined with weighted energy estimates to control the curvature components. By integrating suitable contractions of the Bianchi equations and exploiting the null structure, one obtains energy inequalities that relate fluxes of curvature along null hypersurfaces to spacetime error terms. The use of $r^p$--weighted energies further refines this analysis by emphasizing the contributions of different regions of spacetime and improving the control of decay along outgoing null directions.

We begin by formulating an abstract version of the $r^p$--weighted divergence identities satisfied by pairs of curvature components linked through the null Bianchi equations. 
This formulation isolates the common algebraic structure underlying all curvature components and makes it clear how the weights interact with the transport coefficients in the system.
\begin{df}\label{tensorfields}
For tensorfields defined on a $2$--sphere $(S,\slg)$, we denote by $\sk_0:=\sk_0(S)$ the set of pairs of scalar functions, $\sk_1:=\sk_1(S)$ the set of $1$--forms and $\sk_2:=\sk_2(S)$ the set of symmetric traceless $2$--tensors. This notation reflects the fact that the various null curvature components naturally take values in these spaces.
\end{df}
The following lemma provides the fundamental $r^p$--weighted divergence identity.
It shows that any Bianchi pair gives rise to a spacetime identity whose bulk structure is governed by the transport coefficients $a_{(1)},a_{(2)}$ and the choice of weight $p$. See Lemma 4.2 in \cite{Shen22} for a proof.
\begin{lem}\label{keypoint}
Let $k=1,2$ and let $a_{(1)},a_{(2)}\in\RRR$.
\begin{enumerate}
\item Assume that $\psi_{(1)},h_{(1)}\in\sk_k$ and $\psi_{(2)},h_{(2)}\in\sk_{k-1}$ satisfy the Bianchi-type system
\begin{align}
\begin{split}\label{bianchi1}
\nabs_3\psi_{(1)}+a_{(1)}\trchb\,\psi_{(1)}&=-k\sld_k^*(\psi_{(2)})+h_{(1)},\\
\nabs_4\psi_{(2)}+a_{(2)}\trch\,\psi_{(2)}&=\sld_k(\psi_{(1)})+h_{(2)}.
\end{split}
\end{align}
Then, for any $p\in\RRR$, the pair $(\psi_{(1)},\psi_{(2)})$ satisfies
\begin{align}
\begin{split}\label{div}
&\bdiv\!\left(r^p |\psi_{(1)}|^2e_3\right)+k\,\bdiv\!\left(r^p|\psi_{(2)}|^2e_4\right)\\
+&\left(2a_{(1)}-1-\frac{p}{2}\right)r^{p}\trchb|\psi_{(1)}|^2+k\left(2a_{(2)}-1-\frac{p}{2}\right)r^{p}\trch|\psi_{(2)}|^2\\
=&\;2k r^p \sdivs(\psi_{(1)}\cdot\psi_{(2)})+2r^p\psi_{(1)}\cdot h_{(1)}+2kr^p\psi_{(2)}\cdot h_{(2)}+\lot
\end{split}
\end{align}
\item Assume instead that $\psi_{(1)},h_{(1)}\in\sk_{k-1}$ and $\psi_{(2)},h_{(2)}\in\sk_k$ satisfy
\begin{align}
\begin{split}\label{bianchi2}
\nabs_3\psi_{(1)}+a_{(1)}\trchb\,\psi_{(1)}&=\sld_k(\psi_{(2)})+h_{(1)},\\
\nabs_4\psi_{(2)}+a_{(2)}\trch\,\psi_{(2)}&=-k\sld_k^*(\psi_{(1)})+h_{(2)}.
\end{split}
\end{align}
Then, for any $p\in\RRR$, the pair $(\psi_{(1)},\psi_{(2)})$ satisfies
\begin{align}
\begin{split}\label{div2}
&k\,\bdiv\!\left(r^p |\psi_{(1)}|^2e_3\right)+\bdiv\!\left(r^p|\psi_{(2)}|^2e_4\right)\\
+&k\left(2a_{(1)}-1-\frac{p}{2}\right)r^{p}\trchb|\psi_{(1)}|^2+\left(2a_{(2)}-1-\frac{p}{2}\right)r^{p}\trch|\psi_{(2)}|^2\\
=&\;2 r^p\sdivs(\psi_{(1)}\cdot\psi_{(2)})+2kr^p\psi_{(1)}\cdot h_{(1)}+2r^p\psi_{(2)}\cdot h_{(2)}+\lot
\end{split}
\end{align}
\end{enumerate}
\end{lem}
\begin{rk}
The various null curvature components of the Weyl tensor can all be arranged into Bianchi pairs of the above form. In particular:
\begin{itemize}
\item $(\alpha,\beta)$ satisfies \eqref{bianchi1} with $k=2$, $a_{(1)}=\frac12$, $a_{(2)}=2$;
\item $(\beta,(\rho,-\sigma))$ satisfies \eqref{bianchi1} with $k=1$, $a_{(1)}=1$, $a_{(2)}=\frac32$;
\item $((\rho,\sigma),\underline{\beta})$ satisfies \eqref{bianchi2} with $k=1$, $a_{(1)}=\frac32$, $a_{(2)}=1$;
\item $(-\bb,\aa)$ satisfies \eqref{bianchi2} with $k=2$, $a_{(1)}=2$, $a_{(2)}=\frac12$.
\end{itemize}
\end{rk}
Integrating the divergence identities \eqref{div}–\eqref{div2} over spacetime regions bounded by null hypersurfaces and spacelike slices, and applying Stokes' theorem, one obtains the following $r^p$--weighted energy estimates. 
\begin{prop}\label{keyintegral}
Let $(\psi_{(1)},\psi_{(2)})$ be a Bianchi pair. Depending on the signs of the coefficients,
\[
2+p-4a_{(1)}, \qquad 4a_{(2)}-2-p,
\]
the bulk spacetime terms may or may not yield coercive control of both curvature components. This leads to the following cases.
\begin{itemize}
\item If $2+p-4a_{(1)}>0$ and $4a_{(2)}-2-p>0$, then
\begin{align}
\begin{split}\label{caseone}
&\int_{\cuv}r^p|\psi_{(1)}|^2+\int_{\ucuv} r^p|\psi_{(2)}|^2+\int_V\bigl(r^{p-1}|\psi_{(1)}|^2+r^{p-1}|\psi_{(2)}|^2\bigr)\\ 
\les\;&\int_{\Si_0}\bigl(r^p|\psi_{(1)}|^2+r^p|\psi_{(2)}|^2\bigr)+N(\psi,h).
\end{split}
\end{align}
\item If $2+p-4a_{(1)}\le 0$ and $4a_{(2)}-2-p>0$, then
\begin{align}
\begin{split}\label{casethree}
&\int_{\cuv} r^p|\psi_{(1)}|^2+\int_{\ucuv} r^p|\psi_{(2)}|^2+\int_Vr^{p-1}|\psi_{(2)}|^2\\ 
\les\;&\int_{\Si_0}\bigl(r^p|\psi_{(1)}|^2+r^p|\psi_{(2)}|^2\bigr)+\int_Vr^{p-1}|\psi_{(1)}|^2+N(\psi,h).
\end{split}
\end{align}
\end{itemize}
Here, $N(\psi,h)$ denotes nonlinear terms of the form $\psi_{(1)}\cdot h_{(1)}$, $\psi_{(2)}\cdot h_{(2)}$ and $\Ga\c\psi\c\psi$.
\end{prop}
These estimates make explicit how different curvature components satisfy distinct $r^p$--flux inequalities, reflecting their hierarchy in the peeling structure. This abstract formulation will serve as the starting point for all subsequent curvature estimates. In particular, the choice of $p$ must be adapted to each Bianchi pair in order to obtain coercive bulk control. 
\subsubsection{Strategy of the curvature estimates}
We now summarize how the $r^p$--weighted estimates are combined in \cite{Shen23} in order to control all curvature components and recover the expected decay hierarchy. 
The argument reflects the geometric decomposition of spacetime into exterior and interior regions and proceeds in several steps.
\begin{itemize}
\item One first applies the $r^p$--weighted estimates with $-2\leq p\leq s$ to the Bianchi equations in the exterior region $\Ve$. 
In this region, the geometry is sufficiently close to the asymptotic regime $r\gg1$, and the weighted divergence identities directly provide integrated control of all curvature components with the expected decay behavior.
\item Next, one applies the $r^p$--weighted estimates with $0\leq p\leq s$\footnote{The restriction $p\ge0$ is necessary since the weight $r^p$ becomes unbounded near the symmetry axis $\Phi$ when $p<0$.} 
to the Bianchi equations in the interior region $\Vi=\Vie\cup\Vii$.\footnote{Here, we denote $\Vie:=\{p\in \Vi/\, r(p)\geq \ujp(p)\}$ and $\Vii:=\{p\in\Vi/\,r(p)\leq \ujp(p)\}$.} 

In the intermediate region $\Vie$, the desired decay follows from the geometric relation $\underline{u}\les r$, which allows one to convert $r$--weights into $u$--weights. 
However, in the deep interior region $\Vii$, this mechanism no longer suffices. 
There, the mean value method of Dafermos-Rodnianski \cite{Da-Ro} yields only the weaker decay rate $r^{-\frac{3}{2}}u^{-\frac{s}{2}}$ for the curvature components. 
At this stage, one therefore loses one power of $r$ compared to the decay predicted by the peeling hierarchy.
\item To recover this loss, one commutes the Bianchi equations with $\nabs_T$. 
Since $\nabs_T$ commutes with $\nabs_3$, $\nabs_4$, and $\nabs$ at the linear level, the commuted equations for $\nabs_T\R$ retain the same algebraic structure as the original Bianchi system. Moreover, $\nabs_T\R$ is one order more homogeneous than $\R$. 

This improved homogeneity permits the application of the $r^p$--weighted estimates with the larger range $0\le p\le s+2$ to the commuted system, yielding control of $\nabs_T\R$. 
Using again the mean value method, one obtains the improved decay $r^{-\frac{3}{2}}u^{-\frac{s+2}{2}}$ for $\nabs_T\R$ in $\Vii$.
\item Finally, elliptic estimates for Maxwell-type systems are applied on the spheres $S(t,u)$ in order to transfer the improved decay of $\nabs_T\R$ to $\R$ itself. 
This elliptic step precisely recovers the missing factor of $r$, and yields the optimal decay $r^{-\frac{1}{2}}u^{-\frac{s+2}{2}}$ for the curvature components in $\Vii$.
\end{itemize}
In summary, while the $r^p$--weighted estimates provide the fundamental spacetime control, they are not by themselves sufficient to recover the optimal interior decay. The full hierarchy emerges only after incorporating commutation with $\nabs_T$ and elliptic estimates, which together restore the decay rates predicted by the null structure and peeling properties.
\subsubsection{Estimates for nonlinear terms}
We now turn to the nonlinear error terms arising in the curvature energy identities. 
Their treatment is closely tied to the decay hierarchy established above.

Under the bootstrap assumptions, the optimal decay rates expected for the geometric quantities are
\[
\g-\etabf=O\!\left(\frac{\ep}{r^{\frac{s-1}{2}}}\right),\qquad
\Ga=O\!\left(\frac{\ep}{r^{\frac{s+1}{2}}}\right),\qquad
\R=O\!\left(\frac{\ep}{r^{\frac{s+3}{2}}}\right),
\]
where $\etabf$ denotes the Minkowski metric, $\Ga$ the connection coefficients, $\R$ the Weyl curvature components, and $\ep:=\ep_0^{\frac{2}{3}}$ denotes the bootstrap size.

When $s\in(1,3)$, taking $p=s$ in Proposition~\ref{keyintegral}, the most delicate nonlinear contribution has schematic form
\[
\R \c \Ga \c \R .
\]
Using the above decay rates, one estimates
\begin{align}\label{RGaR}
\int_{V_t} r^s |\R\c\Ga\c\R|\les\int_{V_t}
r^s\frac{\ep}{r^{\frac{s+3}{2}}}\frac{\ep}{r^{\frac{s+1}{2}}}\frac{\ep}{r^{\frac{s+3}{2}}}\les\int_{V_t}\frac{\ep^3}{r^{\frac{s+7}{2}}}\les\ep_0^2.
\end{align}
Since $\frac{s+7}{2}>4$ for $s>1$, the spacetime integral is convergent. 
In particular, the nonlinear terms are strictly subleading compared to the linear bulk terms in the $r^s$--weighted energy identity. Consequently, the bootstrap argument closes at the level of the $r^s$--weighted energy flux, and the nonlinear interactions do not obstruct the decay hierarchy.
\subsection{3D elliptic estimates}
A crucial ingredient in the Christodoulou-Klainerman framework is the use of elliptic estimates on the maximal hypersurfaces $\Si_t$. While the curvature flux identities provide spacetime control of the Weyl tensor, they do not, by themselves, recover the control of the second fundamental form $k$. This gap is bridged by a family of elliptic systems satisfied by the second fundamental form $k$ and its null components.
\paragraph{Div-curl structure of $k$.}
On each maximal hypersurface $\Sigma_t$, the second fundamental form satisfies the elliptic system
\begin{align*}
\sdiv k=0, \qquad \curl k=H, \qquad \tr k=0,
\end{align*}
see Chapter 11 of \cite{Ch-Kl}. Thus, $k$ is determined, in terms of lower-order, by a div-curl system with a source given by the magnetic part $H$ of the Weyl curvature.

Decomposing $k$ relative to the sphere foliation $S(t,u)$,
\[
\ka_{AB}=k_{AB},\qquad 
\eps_A=k_{AN},\qquad 
\de=k_{NN},
\]
with $\kah$ the traceless part of $\ka$, one obtains a coupled elliptic--transport system linking $(\eps,\de,\kah)$ to the curvature components. 
Schematically,
\begin{equation}\label{elliptic_schematic}
\sdivs \kah = \nabs\de +\text{curvature}+\text{lower order}.
\end{equation}
These equations allow one to reconstruct spatial derivatives of $k$ from curvature quantities once suitable weighted estimates are available.
\paragraph{The vectorfield $Z=rN$.}
A key geometric observation in \cite{Ch-Kl} is that contracting $k$ with the vectorfield
\begin{equation}\label{dfZ}
Z:=rN
\end{equation}
produces a $1$--form
\begin{equation}\label{dfiZk}
(i_Z k)_i := k_{ij} Z^j
\end{equation}
which satisfies a div--curl system of the form
\begin{align*}
\sdiv(i_Z k)&=\frac12 \pih^{ij} k_{ij},\\
\curl(i_Z k)&= H_{ij} Z^j + \lot
\end{align*}
Here, $\pih$ denotes the traceless part of the deformation tensor of $Z$. 
This reformulation is crucial: it converts the control of curvature fluxes into a weighted div--curl problem for $i_Z k$.

\paragraph{Weighted elliptic inequality.}
To estimate such systems, one uses a weighted div--curl inequality. 
If a $1$--form $\xi$ satisfies
\[
\sdiv \xi = D(\xi), \qquad \curl \xi = A(\xi),
\]
then for $s\in(1,3)$ one has
\begin{equation}\label{rpdivcurl-est}
\int_{\Sigma_t} \Bigl( r^{s-2} |\nab \xi|^2 + r^{s-4} |\xi|^2 \Bigr)\les\int_{\Sigma_t} r^{s-2}\Bigl( |A(\xi)|^2 + |D(\xi)|^2 + |\sRic| |\xi|^2 \Bigr).
\end{equation}
This is obtained by applying the standard div-curl identity to the weighted quantity $r^{\frac{s-2}{2}} \xi$ and using a Hardy inequality to absorb the lower-order term.
\paragraph{Recovery of decay for $k$.}

Applying \eqref{rpdivcurl-est} to $\xi=i_Z k$ and using the curvature bounds obtained from the $r^p$--weighted energy identities, one obtains
\[
\int_{\Si_t} r^{s-2} |\nab (i_Z k)|^2 +r^{s-4} |i_Z k|^2\les\ep_0^2.
\]
Combined with the definition \eqref{dfiZk}, this yields
\[
\int_{\Si_t} r^{s-2} |(\eps,\de)|^2 \les\ep_0^2.
\]
Finally, the second equation in \eqref{elliptic_schematic} implies the corresponding estimate for $\kah$, and therefore
\begin{align*}
    \int_{\Si_t} r^{s-2} |k|^2 \les \ep_0^2.
\end{align*}
In summary, the 3D elliptic theory on maximal hypersurfaces plays a structural role in the stability argument: it converts curvature control into quantitative bounds for the second fundamental form, restores the decay that is not directly accessible from spacetime energy estimates in the interior region, and closes the hierarchy between curvature and connection coefficients.
\subsection{Exterior stability with borderline decay}
We now explain how exterior stability can still be established in the borderline case $s=1$, as shown in \cite{Shen24}. At the linear level, the decay estimates are essentially the same as in the regime $s>1$. The new difficulty appears in the treatment of nonlinear interactions. Indeed, when $s=1$, repeating the nonlinear estimate \eqref{RGaR} formally yields
\[
\int_{V_t} \frac{\ep^3}{r^4}=\infty,
\]
since in $4$--dimensional spacetime, the measure contributes an additional factor $r^2$ in spherical coordinates. Thus, although the linear decay remains valid, the nonlinear error terms fail to be integrable at the critical decay rate. The obstruction is therefore not a breakdown of linear estimates, but a loss of spacetime integrability for the nonlinear terms.

To recover integrability, one must work with a slightly subcritical weight in the $r^p$--method, namely
\[
p=1-2\de,\qquad 0<\de\ll 1,
\]
in Proposition \ref{keyintegral}. This restores spacetime integrability but leads to weakened decay rates in the exterior region, schematically of the form
\[
\Ga=O\!\left(\frac{\ep_0}{r^{1-\de}\ujp^\de}\right),\qquad
\R=O\!\left(\frac{\ep_0}{r^{2-\de}\ujp^\de}\right).
\]
The loss of the factor $r^{\de}$ reflects the borderline nature of the problem: the $r^p$--weighted estimates alone provide only subcritical control.

Recovering the missing $r^{\de}$ factor, and thus restoring the optimal decay, requires an additional mechanism. In \cite{Shen24}, this is achieved by exploiting transport equations in the incoming null direction. These transport equations propagate improved decay from the asymptotic region inward and compensate precisely for the loss introduced at the level of the $r^p$--weighted estimates. This mechanism also explains the difference in regularity assumptions between \cite{Shen23} and \cite{Shen24}. While $(s,2)$--asymptotically flat initial data suffice when $s>1$, the borderline case $s=1$ requires $(1,3)$--asymptotically flat initial data to control the additional derivatives needed in the transport argument.

In summary, the borderline difficulty is not caused by a failure of linear decay but by the loss of nonlinear spacetime integrability at the critical rate. The $r^p$--weighted method yields only subcritical decay, and the critical loss can be recovered only by exploiting incoming transport equations.
\section{Global stability of Minkowski with borderline decay}
The decay assumption $s>1$ in Theorem \ref{thm:Shen23} is widely believed to be essentially sharp within the framework of current methods. A central open problem is whether the global nonlinear stability of Minkowski spacetime continues to hold in the borderline case $s=1$.

The result of \cite{Shen24} shows that, at the borderline decay level, one can still establish \emph{exterior} stability for $(1,3)$--asymptotically flat initial data in the sense of \eqref{gks1}. However, this leaves open the interior problem and does not address the global geodesic completeness of the full spacetime development. At the critical decay level, the nonlinear analysis becomes genuinely delicate, and the known techniques do not suffice to close the argument globally.

Moreover, one cannot exclude the possibility that $(1,q)$--asymptotically flat initial data sets, for some $q\in\NNN$, may satisfy the borderline decay assumption yet fail to produce a globally stable spacetime. In this direction, the recent work \cite{SW2602} constructs, for any $s\in[1,3)$ and any $q\in\NNN$, nontrivial $(1,q)$--asymptotically flat vacuum initial data which are not $(s,q)$--asymptotically flat, in the sense of Definition \ref{def6.3}, for any $s>1$. In particular, these examples satisfy the hypotheses of \cite{Shen24}, but lie strictly outside the class of initial data covered by \cite{Shen23}. Thus, the borderline case cannot be treated simply as a limiting instance of subcritical theory.

Currently, two fundamentally different scenarios appear conceivable. On the one hand, global nonlinear stability might still hold at the borderline level $s=1$, but its proof would likely require ideas that go substantially beyond the currently available techniques, including the classical vectorfield method, the $r^p$--weighted hierarchy, and the elliptic framework. In particular, new mechanisms would be needed to control borderline nonlinear interactions and to compensate for the loss of spacetime integrability that occurs at critical decay.

On the other hand, it is equally plausible that the critical decay level $s=1$ allows genuinely new dynamical phenomena that are ruled out in the subcritical regime $s>1$. For example, one cannot exclude the possibility that nonlinear effects accumulate in such a way as to produce trapped surfaces in the spirit of \cite{Chr,kr}, or lead to other mechanisms causing a breakdown of global stability.

Determining which of these possibilities is realized, or whether the true behavior lies in a more subtle intermediate regime, remains a central open problem. A definitive resolution would require a deeper understanding of the fine nonlinear structure of the Einstein vacuum equations and may ultimately clarify whether $s=1$ represents a genuine threshold between global dispersive dynamics and qualitatively different long-time behavior.

The resolution of this borderline problem would likely clarify whether dispersive mechanisms alone suffice to prevent singularity formation at critical decay, or whether additional geometric rigidity must be imposed.
\section*{Acknowledgments}
The author is grateful to Lars Andersson and Sergiu Klainerman for the invitation to contribute to this volume. He also thanks Xinliang An, Xuantao Chen, Taoran He, 
Jacques Smulevici, J\'er\'emie Szeftel and Jingbo Wan for valuable discussions on the nonlinear stability problem of Minkowski spacetime which have influenced parts of this survey.

Dawei Shen: Department of Mathematics, Columbia University, New York, NY, 10027. \\
Email: \textit{ds4350@columbia.edu}
\end{document}